\documentstyle[12pt,epsfig]{article}

\catcode`\@=11
\long\def\@makefntext#1{
\protect\noindent \hbox to 3.2pt {\hskip-.9pt
$^{{\ninerm\@thefnmark}}$\hfil}#1\hfill}                

\def\@makefnmark{\hbox to 0pt{$^{\@thefnmark}$\hss}}  

\def\ps@myheadings{\let\@mkboth\@gobbletwo
\def\@oddhead{\hbox{}
\rightmark\hfil\ninerm\thepage}
\def\@oddfoot{}\def\@evenhead{\ninerm\thepage\hfil
\leftmark\hbox{}}\def\@evenfoot{}
\def\sectionmark##1{}\def\subsectionmark##1{}}

\renewcommand{\thefootnote}{\fnsymbol{footnote}}

\def\sectionc{\@startsection {section}{1}{\z@}{-3.5ex plus -1ex minus 
    -.2ex}{2.3ex plus .2ex}{\bf }}
\def\subsectionc{\@startsection{subsection}{2}{\z@}{-3.25ex plus -1ex minus 
   -.2ex}{1.5ex plus .2ex}{\it }}
\renewcommand{\section}[1]{\sectionc{#1}\hspace*{\parindent}}
\renewcommand{\subsection}[1]{\subsectionc{#1}\hspace*{\parindent}}
\newcounter{appendixc}
\newcounter{subappendixc}[appendixc]
\newcounter{subsubappendixc}[subappendixc]

\renewcommand{\appendix}[1] {\vspace*{0.6cm}
        \refstepcounter{appendixc}
        \setcounter{figure}{0}
        \setcounter{table}{0}
        \setcounter{equation}{0}
        \renewcommand{\thefigure}{\Alph{appendixc}.\arabic{figure}}
        \renewcommand{\thetable}{\Alph{appendixc}.\arabic{table}}
        \renewcommand{\theappendixc}{\Alph{appendixc}}
        \renewcommand{\theequation}{\Alph{appendixc}.\arabic{equation}}
        \noindent{\bf Appendix \theappendixc #1}\par\vspace*{0.4cm}}

\def\abstracts#1{{
        \centering{\begin{minipage}{13.2truecm}\footnotesize\baselineskip=13pt\noindent
        \parindent=0pt #1
        \end{minipage}}\par}}

\renewenvironment{thebibliography}[1]
        {\begin{list}{\arabic{enumi}.}
        {\usecounter{enumi}\setlength{\parsep}{0pt}
\setlength{\leftmargin 0.75cm}{\rightmargin 0pt}
         \setlength{\itemsep}{0pt} \settowidth
        {\labelwidth}{#1.}\sloppy}}{\end{list}}

\newcounter{itemlistc}
\newcounter{romanlistc}
\newcounter{alphlistc}
\newcounter{arabiclistc}

\newcommand{\fcaption}[1]{
        \refstepcounter{figure}
        \setbox\@tempboxa = \hbox{\footnotesize Figure~\thefigure. #1}
        \ifdim \wd\@tempboxa > 6in
           {\begin{center}
        \parbox{6in}{\footnotesize\baselineskip=13pt Figure~\thefigure. #1}
            \end{center}}
        \else
             {\begin{center}
             {\footnotesize Figure~\thefigure. #1}
              \end{center}}
        \fi}

\newcommand{\tcaption}[1]{
        \refstepcounter{table}
        \setbox\@tempboxa = \hbox{\footnotesize Table~\thetable. #1}
        \ifdim \wd\@tempboxa > 6in
           {\begin{center}
        \parbox{6in}{\footnotesize\baselineskip=13pt Table~\thetable. #1}
            \end{center}}
        \else
             {\begin{center}
             {\footnotesize Table~\thetable. #1}
              \end{center}}
        \fi}

\def\@citex[#1]#2{\if@filesw\immediate\write\@auxout
        {\string\citation{#2}}\fi
\def\@citea{}\@cite{\@for\@citeb:=#2\do
        {\@citea\def\@citea{,}\@ifundefined
        {b@\@citeb}{{\bf ?}\@warning
        {Citation `\@citeb' on page \thepage \space undefined}}
        {\csname b@\@citeb\endcsname}}}{#1}}

\newif\if@cghi
\def\cite{\@cghitrue\@ifnextchar [{\@tempswatrue
        \@citex}{\@tempswafalse\@citex[]}}
\def\citelow{\@cghifalse\@ifnextchar [{\@tempswatrue
        \@citex}{\@tempswafalse\@citex[]}}
\def\@cite#1#2{{$\null^{#1}$\if@tempswa\typeout
        {IJCGA warning: optional citation argument
        ignored: `#2'} \fi}}

 1
 1
 1

\font\ninerm=cmr9

\begin{document}
\vspace*{-2.5cm}
\begin{flushright}
ADP-96-32/T231\hspace*{2cm}\\
\vspace{1cm}
\end{flushright}
\centerline{\normalsize\bf QUARK DEGREES OF 
FREEDOM IN FINITE NUCLEI\footnote{This work is supported in part
by the Australian Research Council.}}
\baselineskip=15pt

\vspace*{0.6cm}
\centerline{\footnotesize KAZUO TSUSHIMA$^1$,
KOICHI SAITO$^2$ and ANTHONY W. THOMAS$^3$}
\baselineskip=13pt
\centerline{\footnotesize\it $^{1,3}$Department of Physics 
and Mathematical Physics 
and Institute for Theoretical Physics}
\baselineskip=13pt
\centerline{\footnotesize\it University of Adelaide, 
Adelaide, SA 5005, Australia}
\baselineskip=13pt
\centerline{\footnotesize\it $^2$Physics Division, Tohoku College of Pharmacy,
Sendai 981, Japan}
\centerline{\footnotesize E-mail: ktsushim@physics.adelaide.edu.au$^1$}
\centerline{\footnotesize \qquad ksaito@nucl.phys.tohoku.ac.jp$^2$}
\centerline{\footnotesize \qquad \quad \,
 athomas@physics.adelaide.edu.au$^3$}
\vspace*{0.3cm}

\vspace*{0.6cm}
\abstracts{
Properties of finite nuclei are investigated based on
relativistic Hartree equations which have been derived
from a relativistic quark model of the structure of bound nucleons.
Nucleons are assumed to interact through
the (self-consistent) exchange of scalar ($\sigma$) and vector
($\omega$ and $\rho$) mesons at the quark level.
The coupling constants and the mass of the $\sigma$-meson are determined
from the properties of
symmetric nuclear matter and the rms charge radius in $^{40}$Ca.
Calculated properties of static, closed-shell nuclei,
as well as symmetric nuclear matter
are compared with experimental data and with the results
of Quantum Hadrodynamics (QHD).}
%
\normalsize\baselineskip=15pt
\setcounter{footnote}{0}
\renewcommand{\thefootnote}{\alph{footnote}}
\vspace*{0.6cm}

There is no doubt now that explicit quark degrees of freedom for nucleon 
structure are certainly required to understand deep-inelastic scattering 
at momentum transfers of several GeV~\cite{dis}. 
Furthermore, it has also proven possible to successfully 
describe the properties of nuclear matter by taking account of
the quark structure of nucleons~\cite{qmc0,matter}.
Here, we address the question: 
`Are the quark degrees of freedom necessary to describe the properties of
finite nuclei ?', where the typical energy scale is a few tens 
of MeV. Corresponding to this question, we will report our 
recent work on the properties of
finite nuclei (as well as symmetric nuclear matter) 
based on the quark meson coupling 
model (QMC)\footnote{Recently, Blunden and Miller have also
studied the properties of finite nuclei with QMC~\cite{qmc3}.}, 
whose original version 
was suggested by Guichon~\cite{qmc0}.
The main feature of the QMC model is that 
nucleons in the nucleus (matter) are described by the non-overlapping 
MIT bag model within Born-Oppenheimer approximation.

In QMC, the Lagrangian density for mesons and these nucleons can be defined: 
\begin{equation}
{\cal L}=\overline{\psi} [ i \gamma \cdot \partial
-M_N^{\star}({\hat{\sigma}})
-g_\omega\hat{\omega}^\mu \gamma_\mu ] \psi+{\cal L}_{mesons},
\end{equation}
\begin{equation}
{\cal L}_{mesons}=\frac{1}{2}(\partial_\mu\hat{\sigma}\partial^\mu
\hat{\sigma}-
m_{\sigma}^2\hat{\sigma}^2)-\frac{1}{2}\partial_\mu\hat{\omega}_\nu
(\partial^\mu\hat{\omega}^\nu-\partial^\nu\hat{\omega}^\mu)+
\frac{1}{2}m_\omega^2\hat{\omega}^\mu\hat{\omega}_\mu, 
\end{equation}
where, $\psi\ (M_N^\star({\hat{\sigma}}))$, $\hat{\sigma}\ (m_\sigma)$, 
$\hat{\omega}\ (m_\omega)$ are the field operators (masses)
of the nucleon, $\sigma$-, and $\omega$-mesons, respectively, where 
the effective nucleon mass  
$M_N^{\star}({\hat{\sigma}})$ will be defined below.
In mean field approximation, equations of motions for these fields
are given by:
\begin{equation}
[i\gamma \cdot \partial -M_N^{\star}(\sigma)-
g_\omega \gamma_0 \omega ] \psi = 0,\label{eqdirac}
\end{equation}
\begin{eqnarray}
(-\nabla^2_r+m^2_\sigma)\sigma(\vec{r})&=& 
- [\frac{\partial}{\partial \sigma}M_N^{\star}(\sigma)]
\rho_s({\vec r})  = g_\sigma C(\sigma) \rho_s({\vec r}),\label{eqsigma}\\
(-\nabla^2_r+m^2_\omega) \omega(\vec{r}) &=&
g_\omega \rho_B({\vec r}).\label{eqomega}
\end{eqnarray}
On the right hand side of Eq (\ref{eqsigma}), 
a new, and characteristic feature of QMC beyond QHD~\cite{qhd}
appears, namely,  
$- [\frac{\partial}{\partial \sigma}M_N^{\star}(\sigma)]$, or
$C(\sigma)$. These quantities are defined by,
\begin{equation}
\frac{\partial M_N^{\star}}{\partial \sigma}
= -3g_{\sigma}^q \int_{bag} d{\vec r} \ {\overline \psi}_q \psi_q
\equiv -3g_{\sigma}^q S(\sigma) = - \frac{\partial}{\partial \sigma}
\left[ g_\sigma(\sigma) \sigma \right],
\end{equation}
with the MIT bag model quantities
\begin{eqnarray}
M_N^{\star}(\sigma) &=& 
\frac{3\Omega(\sigma(\vec{r}))-z_0}{R_B^{\star}} 
+ \frac{4}{3}\pi ({R_B^{\star}})^3 B ,\nonumber \\
S(\sigma(\vec{r})) &=& \frac{\Omega/2 + m_q^{\star}R_B^{\star}(\Omega-1)}
{\Omega(\Omega-1) + m_q^{\star}R_B^{\star}/2}, \qquad
\Omega = \sqrt{x^2 + (R_B^{\star}m_q^{\star})^2},\nonumber\\
m_q^{\star} &=& m_q - g_{\sigma}^q \sigma (\vec{r}),\quad
C(\sigma) = S(\sigma)/S(0),\quad
g_{\sigma} = 3g_{\sigma}^q S(0).
\end{eqnarray}
Here, $z_0$, $B$, $x$ and $m_q$ are the parameters for the sum of the c.m. 
and gluon fluctuation effects, 
bag pressure, lowest eigenvalue and 
current quark mass, respectively.
$z_0$ and $B$ are fixed by fitting the nucleon mass
in free space, and assumed to be independent of density,  
or $\sigma$-meson field.
The bag radius in-medium, $R_B^\star$, is obtained 
by the equilibrium condition
$d M_N^{\star}(\sigma(\vec{r}))/{d R_B}|_{R_B=R_B^{\star}} = 0$. 
The results reported in this article are obtained with the 
values, $z_0 = 3.295$, $R_B = 0.8$ fm (in free space), 
$B = (170$MeV)$^4$ and 
$m_q = 5$ MeV, respectively.
At the hadron level, the entire information on 
the quark dynamics is condensed
in $C(\sigma)$ of Eq. (\ref{eqsigma}).
Furthermore, when this $C(\sigma) = 1$, the equations of motions 
given by Eqs. (\ref{eqdirac}), (\ref{eqsigma}) and (\ref{eqomega}) 
are exactly identical to those derived from QHD~\cite{qhd}. 
By solving these equations of motions at the 
hadron level, we can investigate the properties of finite nuclei.

\begin{figure}[t]
\begin{center}
\epsfig{file=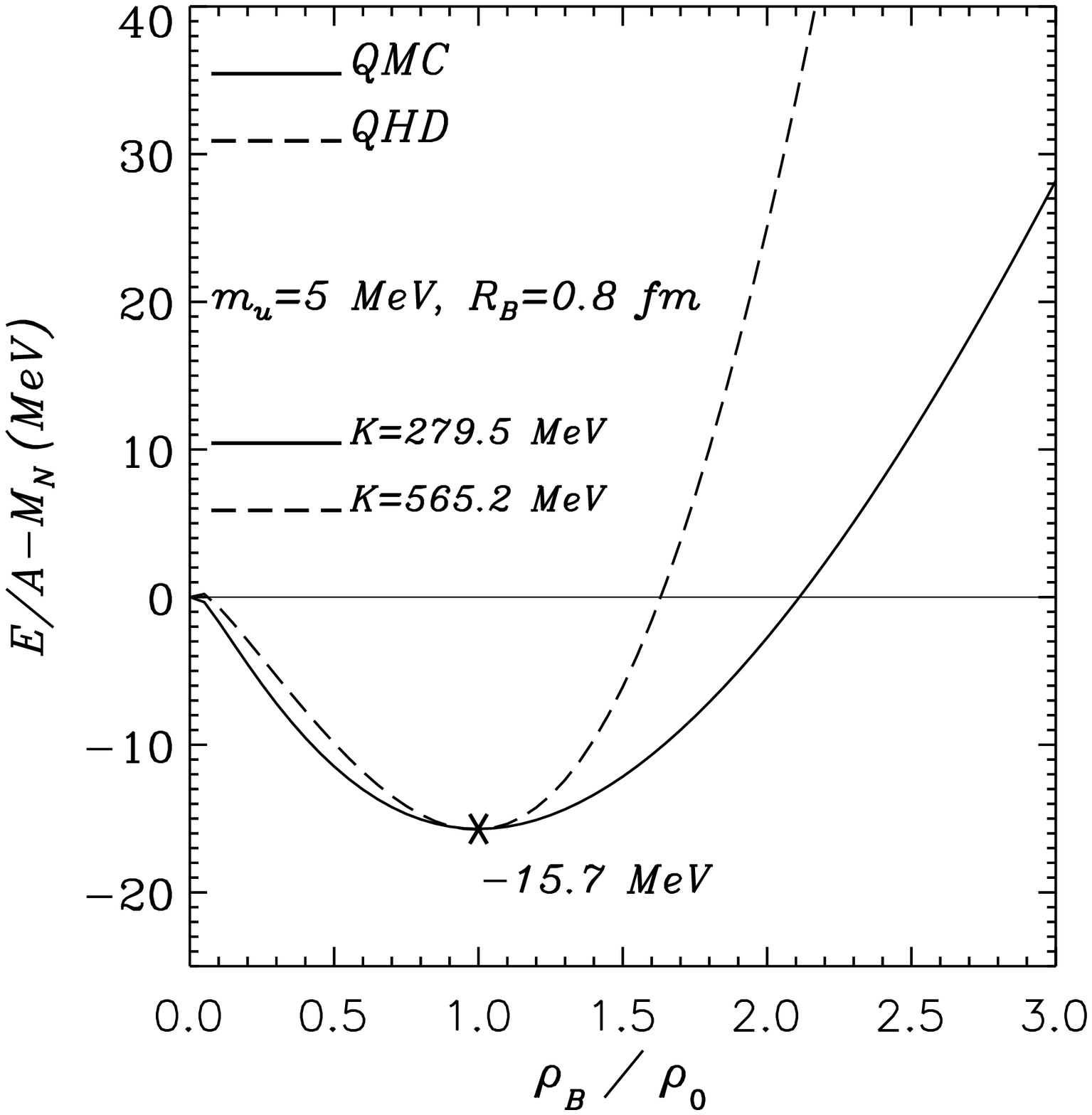,height=7cm}
\end{center}
\fcaption{Calculated (binding energy)/nucleon by QMC (the solid line)
and QHD (the dashed line).
\label{eqs}}
\vspace{-0.5cm}
\end{figure}

We present the calulated results in the following.
For realistic calculations of finite nuclei, contributions 
of the Coulomb force
and the $\rho$ meson are also included~\cite{qmc1,qmc2}.
The model parameters at the hadron level, i.e. coupling constants and mass of 
the sigma meson, $m_\sigma$, are determined from the properties of
symmetric nuclear matter (binding energy per nucleon of -15.7 MeV) and 
rms charge radius in $^{40}$Ca (3.48 fm), with the raitio $g_\sigma/m_\sigma$
fixed so as to reproduce the symmetric nuclear matter 
properties (See Table~\ref{parameter}). 
Other parameters used for the calculations are, 
$m_\omega = 783$ MeV, $m_\rho = 770$ MeV and $e^2/4 \pi = 1/137.036$.
\begin{table}[t]
\protect
\tcaption{Coupling constants and calculated properties 
for symmetric nuclear matter 
at normal nuclear density and finite nuclei.
The effective nucleon mass, $M_N^{\star}$, and the nuclear
compressibility, $K$, are the values for symmetric nuclear matter
at normal nuclear density $\rho_0 = 0.15$ fm$^{-3}$.
$M_N^{\star}$, $K$, and the sigma meson mass, $m_\sigma$, are quoted in MeV.
}\label{parameter}
\small
\vspace{0.4cm}
\begin{center}
\begin{tabular}[t]{c|ccccc|cccc}
\hline
&\multicolumn{5}{c|}{symmetric nuclear matter}&
\multicolumn{4}{c}{finite nuclei}\\
&$M_N^{\star}$&$K$&$m_{\sigma}$&$g_{\sigma}^2/4\pi$&$g_{\omega}^2/4\pi$&
$m_\sigma$&$g_{\sigma}^2/4\pi$&$g_{\omega}^2/4\pi$&$g_{\rho}^2/4\pi$\\
\hline
QMC & 754 & 280 & 550 & 5.40 & 5.31 & 418 & 3.12 & 5.31 & 6.93 \\
QHD & 504 & 565 & 520 & 8.72 & 15.2 & 520 & 8.72 & 15.2 & 5.19 \\
\hline
\end{tabular}
\end{center}
\end{table}
%
The calculated binding energy with results of 
QHD are shown in Fig.~\ref{eqs}.
One of the successes of QMC is that
the nuclear compressibility, $K$, is well reproduced 
the experimentally required values
200 - 300 MeV, whereas QHD tends to overestimate it significantly.
%

\begin{figure}[t]
\begin{center}
\epsfig{file=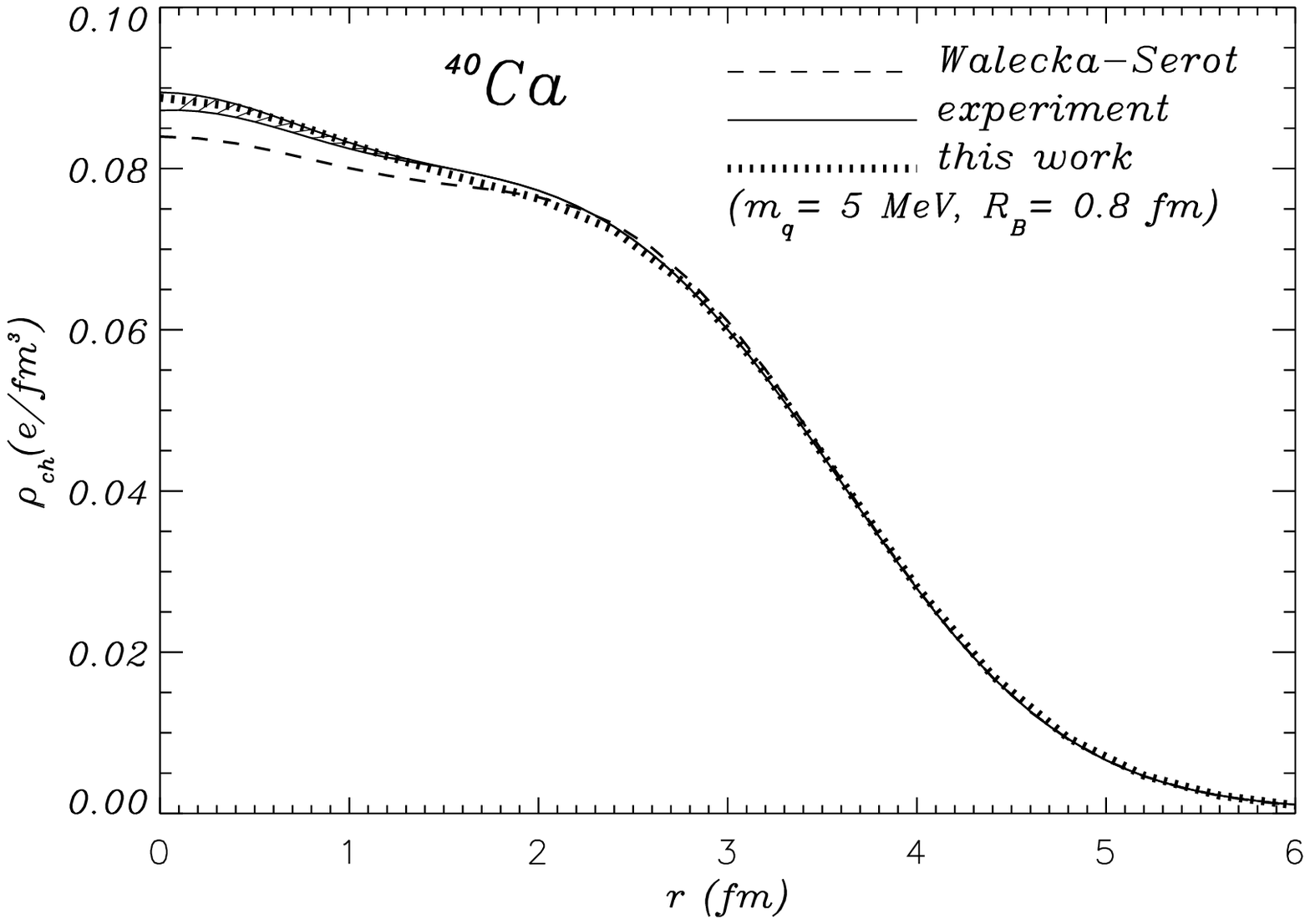,height=6cm}
\end{center}
\fcaption{Charge density distribution for $^{40}$Ca (for $m_q$ = 5 MeV and
$R_B$ = 0.8 fm) compared with the experimental 
data~\protect\cite{expca} and that of QHD.
\label{chca}}
\end{figure}

\begin{figure}[t]
\begin{center}
\epsfig{file=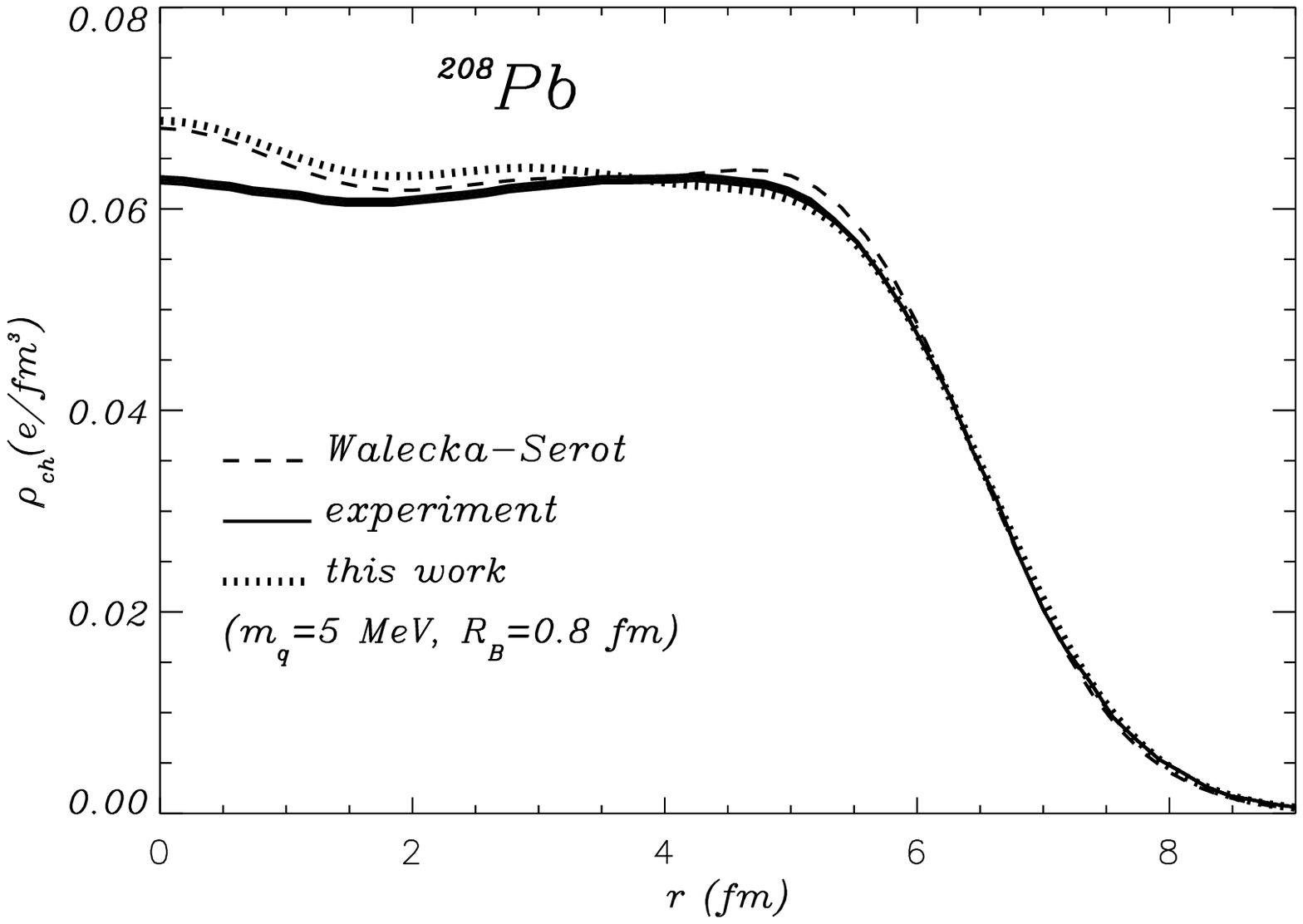,height=6cm}
\end{center}
\fcaption{Charge density distribution for $^{208}$Pb (for $m_q$ = 5 MeV and
$R_B$ = 0.8 fm) compared with the experimental
data~\protect\cite{exppb} and that of QHD.
\label{chpb}}
\end{figure}

In Figs.~\ref{chca} and~\ref{chpb}, we show the calculated charge density
distributions for $^{40}$Ca and $^{208}$Pb with the results of QHD 
and experimental data~\cite{expca,exppb}. 
The charge density is calculated as a
convolution of the point-proton density with
the proton charge distribution~\cite{qmc1,qmc2}.
They are fairly well reproduced, and especially the QMC results for
$^{40}$Ca are impressive. These quantities are not sensitive to
the values of $R_B$ and $m_q$~\cite{qmc1,qmc2}.
%
%

As a summary, we would like to stress the successful generalization 
of the QMC model to finite nuclei
opens a tremendous number of opportunities for further work.
For example, to investigate the Okamoto-Nolen-Schiffer anomaly, 
the nuclear EMC effect, super-allowed Fermi beta-decay, and so on,
which could be resolved by the introduction of the quark degrees of freedom.
Although there are a number of important ways in which this model could be
extended, the present model can be applied to all the
problems for which QHD has proven so attractive, with very little extra
effort. Finally, our answer to the question: 
`Are the quark degrees of freedom necessary to describe the properties of
finite nuclei ?' is that, 
a quantitative investigation has just now started ! \\

\noindent
\vspace*{0.6cm}
{\bf References}
\vspace{-0.8cm}
\end{document}